\documentclass[aps,prl,twocolumn,superscriptaddress,showpacs,showkeys]{revtex4}
\usepackage{bm,amsmath,graphicx}
\begin{document}

\title{Energy conservation in approximated solutions of multiple beam scattering problems}
\author{S\'ergio L. Morelh\~ao}
\email{morelhao@if.usp.br}
\affiliation{Instituto de F\'{\i}sica, Universidade de S\~ao Paulo, CP 66318, 05315-970 S\~aoPaulo, SP, Brazil}
\author{Luis H. Avanci}
\affiliation{Instituto de F\'{\i}sica, Universidade de S\~ao Paulo, CP 66318, 05315-970 S\~aoPaulo, SP, Brazil}
\author{Stefan Kycia}
\affiliation{Laborat\'orio Nacional de Luz S\'{\i}ncrotron/LNLS, CP 6192, 13084-971 Campinas, SP, Brazil}

\date{\today}
\begin{abstract}
Solutions in the form of series expansion, as the Born approximation, are very useful for describing time-independent scattering of quantum particles. In this work, it is mathematically demonstred that such solutions, when applied to multiple beam scattering phenomena, can lead to energy violation where the number of incident and scattered particles is not preserved. General basic conditions for developing consistent solutions of time-dependent multiple beam scattering problems are outlined.
\end{abstract}


\maketitle

Elastic scattering of quantum particles by extended potentials is a well-known phenomenon in Physics. It is extremely important in Material Science where beams of particles, like photons and electrons, are used to probe the atomic structure of the matter. Such phenomena are describable by scattered waves~\textendash~vectorial electromagnetic waves for photons and scalar wave functions for matter particles~\textendash~written in the form of series expansions, for instance
\begin{equation}
\Psi(\bm{r})=\psi_0(\bm{r})+\psi_1(\bm{r})+\psi_2(\bm{r})+\cdots=\sum_{n=0}^{\infty}\psi_n(\bm{r})
\label{se1}
\end{equation}
as been the probability amplitudes accounting for $n^{th}$-order scattering events. The number of terms is related to the extension of the potential, which increases higher-order scattering probabilities. Eq.~(\ref{se1}) stands for the general form of the Born series approximation for time-independent scattering potentials; and although it is conceptually simple and of practical usage, it is quite limited for describing multiple wave scattering problems. In particular, to describe the excitement of one wave when others are already excited since such a situation would configure time-dependent scattering regime.

The first objective of this article is to analytically demonstrate that phenomenological interpretation of multiple wave scattering based on Eq.~(\ref{se1}) leads to energy violation where the potential scatters more particles than exist in the incident beam. Consequently, the Born series expansion should not be taken as a generally valid form of solution for quantitative and accurate description of this type of phenomenon. This remark does not depend on the number of terms considered in the series expansion. The second objective is to develop a general time-dependent solution of the phenomenon that preserves the number of particles entering and leaving the volume occupied by the scattering potential.
\begin{figure} 
\center
\includegraphics[width=2.8in]{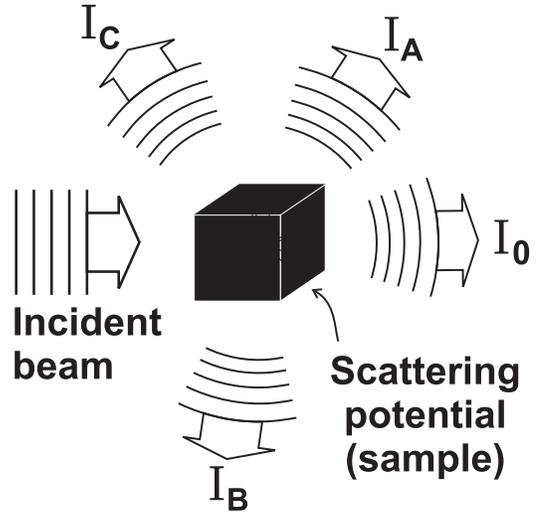} 
\caption{\label{fig1} General configuration of multiple beam elastic scattering of quantum particles (photons, electrons, neutrons, ions,$\cdots$, {\em etc}) by extended potentials (macroscopic samples). It is characterized by scattered waves, $\Psi_{\rm G}(\bm{r})$, within distinct non-overlapping directions of the space. The power of the scattered beams, i.e. the total number of particles per unit of time, are given by $\texttt{I}_{\rm G}=\oint\Psi_{\rm G}(\bm{r})\Psi_{\rm G}^*(\bm{r})\>r^2 {\rm d}\Omega$ (G = 0, A, B, $\cdots$) where the integral is carried out on any closed surface outside the potential's volume.}
\end{figure}

In the standard Born approximation, as find in text books~\cite{cohen}, $\psi_0(\bm{r})$ stands for the incident wave everywhere. It is a reasonable approach for perturbation theories where the scattered beams are much weaker than the incident one. However, for a general treatment considering very strong scattered beams, $\psi_0(\bm{r})$ has to be the incident wave inside the potential's range only; outside, it must give the probability amplitude for non-interacting particles, those passing through the potential without been scattered.

A multiple beam scattering configuration, as schematized in Fig.~\ref{fig1}, occurs when Eq.~(\ref{se1}) can be rewritten as
\begin{equation}
\Psi(\bm{r})=\Psi_0(\bm{r})+\Psi_{\rm A}(\bm{r})+\Psi_{\rm B}(\bm{r})+\cdots
\label{se2}
\end{equation}
since
\begin{equation}
\Psi_{\rm G}(\bm{r})=\sum_{n=0}^{\infty}\psi_{n,{\rm G}}(\bm{r}),\>\>\textrm{G = 0, A, B, $\cdots$ },
\label{se3}
\end{equation}
and
\begin{equation}
\sum_{\rm G}\psi_{n,{\rm G}}(\bm{r})=\psi_n(\bm{r})
\label{se4}
\end{equation}
where $\psi_{0,0}(\bm{r})=\psi_0(\bm{r})$. Since none contribution of the incident wave appears in the directions of the scattered ones, $\psi_{0,{\rm G}\neq0}(\bm{r})=0$; and neither $1^{st}$-order scattering in the direction of the incident wave, $\psi_{1,0}(\bm{r})=0$. Moreover, to characterize a multiple beam scattering configuration, it is also necessary that
\begin{equation}
\oint\Psi(\bm{r})\Psi^*(\bm{r})\>r^2 {\rm d}\Omega=\texttt{I}_0+\texttt{I}_{\rm A}+\texttt{I}_{\rm B}+\cdots = {\rm N}_0
\label{int1}
\end{equation}
where 
\begin{equation}
\texttt{I}_{\rm G}=\oint\Psi_{\rm G}(\bm{r})\Psi_{\rm G}^*(\bm{r})\>r^2 {\rm d}\Omega
\label{int2}
\end{equation}
provides the number of particles per unit of time scattered within distinct non-overlapping solid angles. It guarantees that interference of probability amplitudes occurs exclusively among the series expansion terms of each $\Psi_{\rm G}(\bm{r})$, Eq.~(\ref{se3}). The integrals are carried out on any closed surface outside the potential's volume.

Approximated solution in series expansions are always valid, in principle, for time-independent potentials, or stationary scattering regime. To use the same form of solution for describing the excitement of an extra beam, besides those already excited, further developments are necessary otherwise the incident and scattered number of particles are not preserved in the description of the process. This point is demonstrated here by means of an example. Let us assume a small potential where $3rd$-order scattering terms are negligible, as well as higher order ones, i.e. $\psi_{n\geq3,{\rm G}}\approx0$. According to Born approximation,
\begin{eqnarray}
\Psi_0^{\prime}(\bm{r}) & = & \psi_0^{\prime}(\bm{r})+\psi_{2,0}^{\prime\text{\tiny (A)}}(\bm{r})\nonumber\\
\Psi_{\rm A}^{\prime}(\bm{r}) & = & \psi_{1,{\rm A}}^{\prime}(\bm{r})\label{wv1}
\end{eqnarray}
are the scattered waves under a two-beam excitement condition, $\texttt{I}_0^{\prime}$ and $\texttt{I}_{\rm A}^{\prime}$. On the other hand, when another beam is excited, $\texttt{I}_{\rm B}$ for instance, the scattered waves are given by
\begin{eqnarray}
\Psi_0(\bm{r}) & = & \psi_0(\bm{r})+\psi_{2,0}^{\text{\tiny (A)}}(\bm{r})+\psi_{2,0}^{\text{\tiny (B)}}(\bm{r})\nonumber\\
\Psi_{\rm A}(\bm{r}) & = & \psi_{1,{\rm A}}(\bm{r})+\psi_{2,{\rm A}}^{\text{\tiny (B)}}(\bm{r})\label{wv2} \\ 
\Psi_{\rm B}(\bm{r}) & = & \psi_{1,{\rm B}}(\bm{r})+\psi_{2,{\rm B}}^{\text{\tiny (A)}}(\bm{r})\>\>.\nonumber
\end{eqnarray}
{\small (A,B)} superscripts are used on $2^{nd}$-order waves to identify the $1^{st}$-order ones originating them. For instance, $\psi_{2,0}^{\text{\tiny (A)}}(\bm{r})$ stands for the rescattering of $\psi_{1,{\rm A}}(\bm{r})$ towards the forward-transmitted wave, $\psi_0(\bm{r})$.

The number of particles entering and leaving the effective range of the potential must be preserved under pure elastic scattering. It means that for an incident beam constant in time, N$_0$ in Eq.~(\ref{int1}) is constant, and therefore, the total scattered intensities of the waves in Eqs.~(\ref{wv1})~and~(\ref{wv2}) are the same, i.e. $\texttt{I}_0^{\prime}+\texttt{I}_{\rm A}^{\prime}=\texttt{I}_0+\texttt{I}_{\rm A}+\texttt{I}_{\rm B}$. It leads to
\begin{widetext}
\begin{equation}
\oint\left\{|\psi_0^{\prime}+\psi_{2,0}^{\prime \text{\tiny (A)}}|^2+
|\psi_{1,{\rm A}}^{\prime}|^2 \right\}\>r^2 {\rm d}\Omega=
\oint\left\{|\psi_0+\psi_{2,0}^{\text{\tiny (A)}}+\psi_{2,0}^{\text{\tiny (B)}}|^2+
|\psi_{1,{\rm A}}+\psi_{2,{\rm A}}^{\text{\tiny (B)}}|^2+
|\psi_{1,{\rm B}}+\psi_{2,{\rm B}}^{\text{\tiny (A)}}|^2 \right\}\>r^2 {\rm d}\Omega\>\>.
\label{int3}
\end{equation}
\end{widetext}

The challenge in describing the scattering of multiple beams by approximated solutions can be summarized in the above example, in how to go from Eq.~(\ref{wv1}) to Eq.~(\ref{wv2}) without violating the equality in Eq.~(\ref{int3}). In more specific words, it is necessary to described not only how the extra terms in Eq.~(\ref{wv2}) are excited, but also how the terms already excited in Eq.~(\ref{wv1}) are affected by the excitement of the new beam, $\texttt{I}_{\rm B}$ in this case. For a qualitative description, one may assume that $\psi_0=\psi_0^{\prime}$, $\psi_{2,0}^{\text{\tiny (A)}}=\psi_{2,0}^{\prime \text{\tiny (A)}}$, and $\psi_{1,{\rm A}}=\psi_{1,{\rm A}}^{\prime}$, and that the extra terms in Eq.~(\ref{wv2}) are switched on as the potential varies in time. However, in this phenomenological description the total number of scattered particles is not preserved; unless for a very particular coincidence where $\psi_{2,{\rm G}}^{\text{\tiny (B)}}$ would provide destructive interference effects with the other terms of $\Psi_{\rm G}(\bm{r})$ (G = 0 or A) so that these destructive interference would account for the exact number of particles in $\texttt{I}_{\rm B}$. Although some phase relationships may exist among the scattered waves, they vary from one potential to another since the phases are intrinsically related to the internal three-dimensional structure of the scattering potentials. In the next part of this article, a general time-dependent solution, in form of series expansion, is proposed for describing the multiple beam scattering phenomenon.

Conservation of the number of scattered particles in approximated solutions is possible by taken into account the scattering probabilities of the particles when traveling through the potential. By introducing $p_{\rm H,G}(n,t)$ as the scattering probability from beam G to beam H after $n$ scattering events, the population of particles in the beam H that have already been scattered (bounced) from one beam to another $n+1$ times is
\begin{equation}
{\rm P}_{\rm H}(n+1)=\sum_{\rm G} p_{\rm H,G}(n,t) {\rm P}_{\rm G}(n)
\label{rec1}
\end{equation}
where $p_{\rm H,G}(n,t)=0$ for H = G, and the time, $t$, dependence of these probabilities are determined by the potential, i.e. on how it varies in time until a multiple beam scattering configuration is achieved. Moreover, assuming a slow time variation to assure that at any time instant the populations are given by Eq.~(\ref{rec1}), the behavior of the populations as a function of $n$ can be inferred by the sum of probabilities,
\begin{equation}
{\rm s}_{\rm G}(n)=\sum_{\rm H} p_{\rm H,G}(n)\>. 
\label{sum1}
\end{equation}
When s$_{\rm G}(n)=1$ the $n$-bounced particles in the beam G have 100\% probability to be scattered towards another beam before leaving the potential's range. Therefore, outside the potential, the total number of particles on beam G is calculated as 
\begin{equation}
\texttt{I}_{\rm G}=\sum_{n} [1-{\rm s}_{\rm G}(n)]{\rm P}_{\rm G}(n)\>.
\label{sum2}
\end{equation}
It implies that when s$_{\rm G}(n)=1$ none $n$-bounced particles are found on beam G. On the other hand, when s$_{\rm G}(n)\ll1$ the potential scatters a significant fraction of the population of $n$-bounced particles through beam G.

A direct comparison between Eqs.~(\ref{int2}) and (\ref{sum2}) leads to a straightforward conclusion that the $\Psi_{\rm G}(\bm{r})$ waves, as given in Eq.~(\ref{se3}), must be redefined outside the potential as
\begin{equation}
\Psi_{\rm G}(\bm{r})=\sum_{n=0}^{\infty}\sqrt{1-{\rm s}_{\rm G}(n)}\>\psi_{n,{\rm G}}(\bm{r})\>. 
\label{se5}
\end{equation}

The physical meaning of $\sqrt{1-{\rm s}_{\rm G}(n)}$ is clear. It reduces the probability amplitudes of scattering towards beam G when other beams are excited, i.e. when the particles on beam G, inside the potential, have a non null probability of leaving the potential's volume via other beams. Let us use the three-beam configuration represented in Eq.~(\ref{wv2}) to demonstrated that the above form of solution, Eq.~(\ref{se5}), can be used to described the excitement of the third beam and at the same time preserving the total number of scattered particles, ${\rm N}_0$.

The small potential responsible for the scattered waves in Eq.~(\ref{wv2}) can be represented by a set of scattering probabilities, so that ${\rm s}_0=p_{\rm A,0}+p_{\rm B,0}$, ${\rm s}_{\rm A}=p_{\rm 0,A}+p_{\rm B,A}$, and ${\rm s}_{\rm B}=p_{\rm 0,B}+p_{\rm A,B}$ are taken as constant values as a function of $n$. Then, by upgrading the set of scattered waves in Eq.~(\ref{wv2}) according to the new definition given in Eq.~(\ref{se5}), and replacing then into Eq.~(\ref{int2}) we obtain
\begin{eqnarray}
\texttt{I}_0&\simeq&{\rm N}_0
(1-p_{\rm A,0}-p_{\rm B,0}+p_{\rm 0,A}p_{\rm A,0}+p_{\rm 0,B}p_{\rm B,0})\nonumber\\
\texttt{I}_{\rm A}&\simeq&{\rm N}_0
(p_{\rm A,0}+p_{\rm A,B}p_{\rm B,0}-p_{\rm 0,A}p_{\rm A,0}-p_{\rm B,A}p_{\rm A,0})\nonumber\\
\texttt{I}_{\rm B}&\simeq&{\rm N}_0
(p_{\rm B,0}+p_{\rm B,A}p_{\rm A,0}-p_{\rm 0,B}p_{\rm B,0}-p_{\rm A,B}p_{\rm B,0})\nonumber\\
\label{su4}
\end{eqnarray}
where triple products of the $p_{\rm H,G}$ probabilities are disregarded since $\psi_{n\geq3,{\rm G}}\approx0$ for such a small potential. 

According to the above intensities, Eqs.~(\ref{su4}), the total number of particles is always preserved, i.e. 
\begin{equation}
\texttt{I}_0+\texttt{I}_{\rm A}+\texttt{I}_{\rm B}\simeq{\rm N}_0\>
\label{su5}
\end{equation}
independently of the number of scattered beams. For instance, when $p_{\rm B,0}=p_{\rm B,A}=0$ beam B is not excited, but the sum of intensities still provide the incident number of particles, $\texttt{I}_0^{\prime}+\texttt{I}_{\rm A}^{\prime}\simeq{\rm N}_0$. If beam A is also switch off, $p_{\rm A,0}=p_{\rm A,B}=0$, no particle-potential interactions occur so that $\texttt{I}_0^{\prime\prime}={\rm N}_0$.

Strong experimental evidences suggesting scattered waves as given in Eq.~(\ref{se5}) are found in recent X-ray multiple diffraction experiments~\cite{more2002}. There, the linear polarization of synchrotron radiation is exploited to enhance the {\em Aufhellung} effect~\cite{wagner,chang}. In terms of scalar wave functions, such an experiment was equivalent of monitoring 
\begin{equation}
\texttt{I}_{\rm A}=\oint(1-{\rm s}_{\rm A})\left|\psi_{1,{\rm A}}+\psi_{2,{\rm A}}^{\text{\tiny
(B)}}\right|^2 \>r^2 {\rm d}\Omega
\label{int4}
\end{equation}
when exciting $\texttt{I}_{\rm B}$. The outcome is a well defined intensity reduction owing to the relative enhancement of ${\rm s}_{\rm A}$ if the contribution of $\psi_{2,{\rm A}}^{\text{\tiny (B)}}$ is eliminated.

In summary, multiple wave scattering configurations are mathematically defined. It demonstrates that the number of incident and scatted particles is not preserved when time-independent solutions in form of series expansion are extended to describe time-dependent scattering problems. Basic requirements of time-dependent solutions are pointed out. It could also be used, in principle, to upgrade available solutions~\cite{shen} to account for the {\em Aufhellung} effect, commonly observed in multiple beam X-ray diffraction.

\begin{acknowledgments}
Financial supports are received from Brazilian founding agencies FAPESP, grant number 02/10387-5, and CNPq, proc. number 301617/95-3.
\end{acknowledgments}


\begin{thebibliography}{5}
\bibitem{cohen} Claude Cohen-Tannoudji, \textit{Quantum Mechanics}, A Wiley-Interscience publication, New York (1977).
\bibitem{more2002} S.L.~Morelh\~ao and S.~Kycia, Phys. Rev. Lett. \textbf{89}(1), 015501 (2002). 
\bibitem{wagner} E.~Wagner, Phyz. Z. \textbf{21}, 94 (1923).
\bibitem{chang} S.-L. Chang, \textit{Multiple Diffraction of X-Rays in Crystals}, Springer Verlag (1984). 
\bibitem{shen} Q. Shen, Acta Cryst. A\textbf{42}, 525 (1986). 
\end{thebibliography}
\end{document}